# Graphene-based Soft Wearable Antennas

[1]‡Isidoro Ibanez-Labiano, [2]M. Said Ergoktas, [2]Coskun Kocabas, [3]Anne Toomey, [1]Akram Alomainy, [3]*Elif Ozden-Yenigun

[1]School of Electronic Engineering and Computer Science, Queen Mary University of London, United Kingdom
[2]Department of Material and National Graphene Institute, University of Manchester, United Kingdom
[3]School of Design, Textiles, Royal College of Art, London, United Kingdom
*elif.ozden-yenigun@rca.ac.uk

## Abstract

Electronic textiles (e-textiles) are about to face tremendous environmental and resource challenges due to the complexity of sorting, the risk to supplies and metal contamination in textile recycling streams. This is because e-textiles are heavily based on the integration of valuable metals, including gold, silver and copper. In the context of exploring sustainable materials in e-textiles, we tested the boundaries of chemical vapour deposition (CVD) grown multi-layer (ML) graphene in wearable communication applications, in which metal assemblies are leading the way in wearable communication. This study attempts to create a soft, textile-based communication interface that does not disrupt tactile comfort and conformity by introducing ML graphene sheets. The antenna design proposed is based on a multidisciplinary approach that merges electromagnetic engineering and material science and integrates graphene, a long-lasting alternative to metal components. The designed antenna covers a wide bandwidth ranging from 3 GHz to 9 GHz, which is a promising solution for a high data rate and efficient communication link. We also described the effects of bending and proximity to the human body on the antenna's overall performance. Overall, the results suggested that graphene-based soft antennas are a viable solution for a fully integrated textile-based communication interface that can replace the current rigid, restrictive and toxic approaches, leading to a future where eco-friendliness and sustainability is the only way forward!

## 1. Introduction

Smart textiles have become an active research field in recent decades, with increasing interest in wearable and flexible platforms for the integration of various functionalities, from sensor-based devices to user-interface elements. Textiles that provide a seamless command-oriented user interface [1] and that are capable of wireless communication [2-3] have been an increasingly popular research topic in recent years. The notion of electronic textiles (e-textiles) has recently shifted toward more active and smarter functions, promoting textile materials and structures with inherent electronic capabilities, such as piezoelectric yarns, photovoltaic fibres and sensory fabrics. As stated by [4] Köhler *et al.*, a combination of relatively short-lived mass-consumer goods, electronics and textiles will form a new type of waste, namely e-waste, including contaminated used textiles. Our experiences with the disposal of contemporary electronic waste give us reason to expect that severe global environmental and social impacts will also result from the recycling and disposal of e-textiles.

Until now, e-textiles have used a similar range of materials to those used in other current commercial electronic products. That is, they all contain a substantial amount of a number of valuable metals such as copper, silver and gold, as well as a range of hazardous substances and their precursors [5]. Thus, e-textiles that function at the intersection of metals and either man-made or natural textile forms combine different extremes of material properties – for example, softness and stiffness, washability and susceptibility to corrosion. Meeting the requirement for textiles to exhibit qualities of drapability, touch, lightness and washability when combined with metals therefore creates new challenges in making the textiles wearable. Researchers are pursuing different design strategies for the integration of electronic components according to their respective disciplines. As with many e-textile applications, including health monitoring in MedTech, wearable computing, battery design and energy-harvesting textile design, in the era of the Internet of things (IoT) smart textiles still rely heavily on the use of metal in wearable communication.

Antennas are a vital point in the design of body-centric wireless systems that require the guarantee of a satisfactory power transfer between nodes placed on different parts of the body, as well as communication with off-body devices [6]. For wearable communications systems, mechanical flexibility and effective radiation also require passive components, including transmission lines and antenna that are used to send and receive radio frequency (RF) signals. Hence, conductive metals that ensure off-the-shelf availability and reliable solutions are leading the field of body-centric communication. In textile-based antennas, methods such as embroidery, printing – including inkjet and screen-printing techniques – and patterning conductive patches have been studied extensively over the last few years. Remarkably, the very early successful prototypes, embroidered antennas [7-15] and microstrip patch antennas [16-18] integrated into textiles, are still employed by tuning the geometries responding to different communication protocols. The use of copper and other conductive foils, and pleated metal threads in making textile-based antennas, built a knowledge of textile uses in communication. That, in turn, has informed a design process based on capturing electric properties and their effect on performance in relation to various parameters, such as the conformity of textiles, which are therefore sensitive to bending, stretching, and compression [19]. The emphasis on creating conformal antenna surfaces specifically for textiles has pushed the exploration of new manufacturing methods. Commonly used printing techniques, including inkjet printing [20] and screen-printing [21], have opened up new making strategies for on-body antenna systems. It is apparent, however, that all these scalable possibilities in wearable antenna fabrication are restrained by the conductive element, which quickly becomes corroded and oxidized, and also brings high material costs due to the loss of performance [22]. However, many studies have continued to focus on different metallic materials, in the form of conductive copper and silver inks, Cu/Ag/Au pleated yarns and Cu/Ag/Au/Al/Ti conductive coatings, in leading the research to improve on traditional bulky antennas. As opposed to many textile substrates, additional measures need to be taken in protecting these metal antennas against water absorption and corrosion.

As an alternative to conventional materials on the market, carbon nanomaterials such as graphene possess various outstanding material properties, including electrical sensitivity, piezoresistivity, high optical transmittance and high strength/stiffness. The foreseen challenges in IoT pointed out the need for flexible and stretchable sensing and communication systems for wearable technologies. In the interest of proposing wearable systems, graphene-based nanostructure assemblies have been successfully employed as flexible gas and chemical sensors [23] and wireless strain and tactile sensors [24-26]. The very same motives are leading the researchers to fabricate flexible radio-frequency identification (RFID) tags by ink-jet printing [27-30]. A pioneering work in field [31], Huang *et al.* designed and fabricated simple printed graphene that enabled transmission lines and antenna structures on paper substrates. In another study, Huang *et al.* revealed that graphene ink could potentially be a low-cost alternative to much more expensive metal nanoparticle inks, such as silver nanoparticle ink [32]. In the move towards graphene-based wearable antenna systems, many other material systems, including graphene nano ribbon-based terahertz patch antenna on polyimide [33], inkjet graphene dipole antenna on cardboard [34], 3D direct written graphene radio-frequency identification (RFID) antennas on textile, wood and cardboard substrates [30] and conductive ink comprising graphene nanoplatelets deposited on cotton fabric [35], transparent dipole antenna [36], microwave antennas [37] have produced low-cost, effective solutions for wireless communications. Recently, Jaakkola *et al.* [38] printed graphene ink-based dipole antennas combined with RFID microchips on Kapton using screen-printing and on paper using spray-coating. Both graphene and graphene oxide inks have their own limitations in terms of electrical conductivity, due to the interrupted conductive surfaces governed by the lateral size of their particles. When coupled with heterogeneous structures such as woven, knitted and nonwoven textiles, screen-printed, inkjet- and 3D-printed antennas are more prone to damage, operation frequency shifts and losses, due to the compressing and tensioning of the coated surface [20].

Recently, Ergoktas *et al.* proposed a unique fabrication method to manufacture infrared textile devices, including display, yarn and stretchable devices, by means of ML graphene lamination [39]. This attachment technique for textiles has opened up new making possibilities that have successfully benefited from the intrinsic merits of CVD grown graphene sheets.

All of these prospective challenges, such as the disruption of the softness and lightness of fabrics that incorporate antennas, have provided the impetus for us to create new material systems that seek a deep understanding of smart ways of using 2D materials in e-textiles. This study has thus attempted to develop soft graphene-based antennas for smart textiles that extend the possibilities of current state-of-the-art wireless body-centric systems. For the first time, to the best of our knowledge, we have introduced CVD grown ML graphene sheets on a cellulosic textile substrate to produce antennas for body-centric wireless communication. The proposed design, consisting of CPW-fed planar inverted cone-shaped patch geometry, is tuned to ensure the wearer's comfort by eliminating the additional buffer layers and stiff components that are often used in radiating and ground layers. The results suggest that graphene-based textile antennas are capable of responding to a broad bandwidth with a slight detuning when placed on the body. The conformity of the antenna structure promoted good adaptation, enabling the graphene-based textile antennas to maintain their responsiveness under bending.

## 2. Materials and Method

### 2.1. Synthesis of Multi-layer Graphene and Transfer Printing of Graphene on Polyethylene Films

ML graphene was synthesized on 25 µm thick Nickel foil (Alfa Aesar #12722) by CVD method. Ni foil was placed in the CVD chamber and heated to 1050$_o$C under 100sccm $H_2$ and 100Ar gas flow, then annealed at the same conditions for 20 minutes. An additional 35sccm $CH_4$ flow was used as a carbon precursor for 15 minutes at atmospheric pressure during the growth stage. Lastly, the sample was cooled down to room temperature rapidly under 100sccm $H_2$, and 100sccm Ar flow. The quality of the synthesized ML graphene samples was determined by Raman spectroscopy by using 532 nm 50 mW laser source and 50x objective, as seen in Figure 1. The Raman spectrum clearly shows that the quality of ML graphene on Ni is good, as there is no D peak detected at ~1350 cm$_{-1}$ (defect peak, D) while G peak at 1580 cm$_{-1}$ exhibits a higher intensity than 2D peak at 2713 cm$_{-1}$. The full width half maxima (FWHM) values were calculated as 20.7cm$_{-1}$ and 66.1cm$_{-1}$ for G and 2D bands, respectively. After the ML graphene synthesis, 20 µm thick polyethylene (PE) was laminated onto ML graphene and Ni was etched in 1M $FeCl_3$ solution for ~8 hours, which was followed by rinsing in deionized (DI) water.

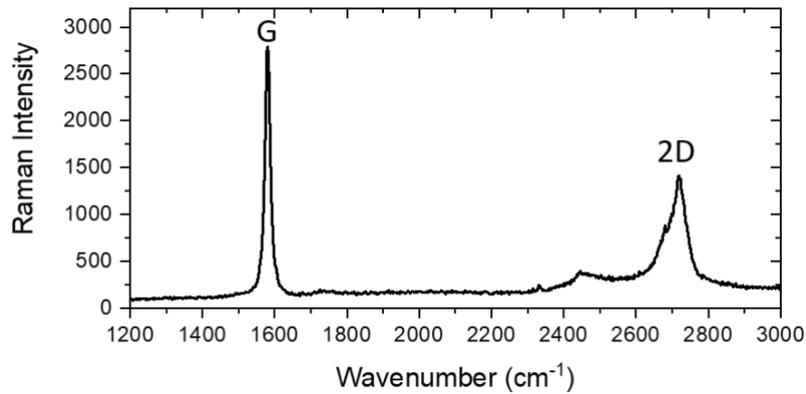

*Figure 1.* Raman spectrum of CVD grown ML graphene by using a 532 nm 50 mW laser source and 50x objective.

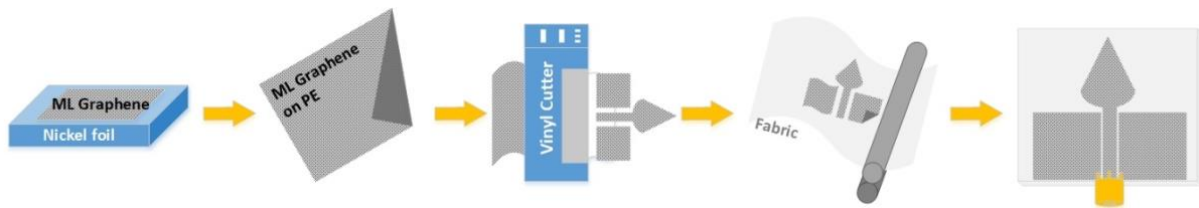

*Figure 2.* The fabrication process of graphene-based soft antennas. From left to right: *(i)* ML graphene grown on Ni foil; *(ii)* ML transferred on PE sheets; *(iii)* cutting the patterns; *(iv)* graphene lamination on textiles, *(v)* soft graphene based antenna with an attached SMA connector.

## 2.2. Fabrication of Graphene-Based Textile Antennas

The emphasis on its ability to be worn, and the antenna's performance when coupled with the human body, revealed an intriguing on-body presentation. In terms of electromagnetic properties, the human body can be considered as a lossy medium associated with high dielectric constant, which in return detunes the antenna by causing frequency shifts and affects the antenna's gain and its efficiency [40-41]. Among standards of communications, ultra-wideband (UWB) technology provides high capacity with large bandwidth signals, multi-path robustness and low-power systems. UWB antennas are ideal candidates for body-worn units in sports analytics [42], health monitoring [43-44] and positioning systems [45]. Thus, the effect of the presence of the human body has been studied several times while its impact on the performance of UWB antennas has been analysed [46]. Nevertheless, locating the antenna on the body depends on several factors, such as the expected performance in such applications. Several studies exploring different positioning of antennas for body-worn systems have pointed to the torso as the most convenient position for combining both on-body and off-body communications [47]. The geometry of the design of the UWB wearable antenna is based on the final application [48] and the communication channel, in both off-body and on-body communication [49]. Within wearable technologies, antennas must be introduced to the system without disrupting the user's performance or comfort. Thus, in a body context wearable antennas must be compact, conformal and low profile, making planar structured designs ideal candidates for these systems. The co-planar waveguide (CPW) technique is a well-established solution for wearable antennas due to its reduced fabrication complexity that avoids the misalignment of layers. This technique also enables easy integration into textiles and facilitates new fabrication methods such as lamination processes or inkjet printing [50-52].

Figure 2 depicts the fabrication process of flexible ML graphene transferred onto textiles. The main challenges in smart material integration arise from the fact that the antenna patches exhibit different conformity and haptic responses from those of textiles. In order to enable full interconnectivity, the geometry of the antenna design and its effects on the propagation channel should be carefully and systematically evaluated. Thus, as illustrated in Figure 2, a very thin adhesive layer of EVA was applied on ML graphene on PE samples to promote binding at the fabric interface. For successful alignment of different components in the design, a vinyl cutter was used to create a paper mould. As a last step of the fabrication, we placed the ML graphene on PE samples on the adhesive with the conductive side facing up, to allow easy contact with antenna sockets/pins from MLG, and laminated it again to provide a strong interface and mechanical stability.

ML graphene samples that were later modelled as the conductive material in antenna measurements were tested with a four-probe method. The sheet resistance of ML graphene on the PE sample was *c.a* 25 Ω/sq. The conductivity of graphene could be tuned at the range of microwave to terahertz (THz) frequencies by an external field or chemical doping. Over a broad range, the conductivity of ML-graphene sheets found independent [53]. Using the proposed method, either bulk or thin-film specimens can be tested by marking the thickness of the substrate.
Sheet resistance ($R_s$) is measured as

$R_s = \rho / t$ *(Equation 1)*

where $\rho$ is the resistivity and $t$ the thickness material.
For a very thin sheet of material, ($R_s$) is

$R_s = k\,(V/I)$ *(Equation 2)*

where the $k$ is a geometric factor. In the case of a semi-infinite thin sheet, k is assumed to be 4.53 [54-55].

One of the significant concerns in smart wearables is the washing durability of conductive components. We performed a washability test by using the standard protocol in BS EN ISO 6330. Both stand-alone ML graphene sheets on PE and ML graphene laminated cotton fabrics were washed in a Type A reference washing machine, by following the procedure no. 4H at 40 ± 3 C, and using a detergent without an optical brightener. The resistivity of the ML graphene sheets and the ML graphene on the cotton fabric were also measured after 50 washing cycles.

After the transfer printing of CVD grown ML graphene onto fabric, silver-based conductive epoxy (SPI Silver Conductive Epoxy 05000-AB) curable at room temperature was used to attach SMA 3-pin 50 Ohm connectors for the feeding. It was then cured for approximately 4 hours at 75°C. The final prototype, as shown in Figure 3, facilitates the performance evaluation of both the graphene-radiating component and the graphene base layer on the front, demonstrating whole-graphene antenna systems for ultra-wideband communication.

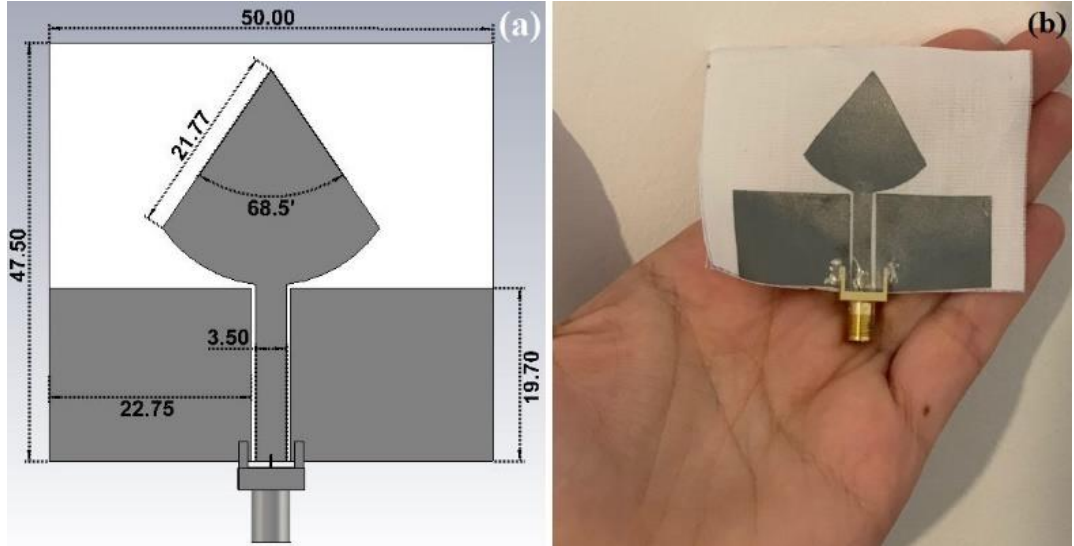

*Figure 3. (a)* PICA antenna design CAD model with the dimensions in mm *(b)* Actual soft graphene antenna prototype.

One of the major concerns associated with wearables is the weight penalty associated with the radiating and base components. As a comparison, we fabricated a reference antenna with copper tape on a cotton fabric substrate, using the geometries in the PICA model [3], which weighed approximately 0.75g. This is 25% heavier than ML graphene-laminated antennas. It is important to note that the adhesive layer and the PE substrate, rather than the graphene sheets themselves, constitute a significant proportion of the weight of the antenna.

## 2.3. Antenna Characterization and Simulations of Graphene-Based Textile Antennas

The modelling of antenna prototypes requires an accurate representation of materials. The electromagnetic properties of a material can be calculated using several different methods. According to microwave theory, they can be classified into two complementary sets: non-resonant and resonant. Resonant methods provide more accurate results than non-resonant methods at a single frequency or at several discrete sets of frequencies. Similarly, resonant methods could also be divided into two sub-categories: resonant perturbation and resonator methods [56-57]. In this study, a resonant perturbation method was used to measure the dielectric properties of the substrate textile materials.

Two parameters addressing the dielectric characteristics of materials, such as the dielectric constant ($\varepsilon_r$) and the dissipation factor ($DF$), were measured. The absolute permittivity of a material ($\varepsilon$), often known as permittivity, describes how a material behaves when subjected to an electric field, and it is represented as the complex number $\varepsilon$

$$\varepsilon = \varepsilon_r \varepsilon_o = (\varepsilon_r' - j\varepsilon_r'') \varepsilon_o \qquad \text{(Equation 3)}$$

where $\varepsilon_o$ is the vacuum permittivity (~8.85419–12 F·m–1) and $\varepsilon_r$ refers to the polarized molecules within the material under an electric field. While the dissipation factor ($DF$), also called the loss tangent $tan\delta$, is a ratio of the loss index and the relative permittivity. It is the inverse of the quality factor ($Q$), and is calculated by the following equation:

$$tan\delta = \varepsilon_r'' / \varepsilon_r' = DF = 1/Q \qquad \text{(Equation 4)}$$

To derive the textile's dielectric properties, dielectric constant and dissipation factor, from the existing range of resonant methods the perturbation cavity method was employed. The fabric sample was placed inside an aperture of the cavity, in which the reflected waves and the transmitted waves at the interface between the fabric and the cavity were monitored in order to determine the dielectric properties of the textile under characterization [58]. For this specific purpose, a split cylinder resonator for material characterization (Agilent 85072A), working at 10 GHz, and the Keysight material characterization software (N1500A-003 Materials Measurement Suite 2015) were used. This single frequency characterization method that ensures good control on cavity specifications and boundary conditions, provides more accurate representation than other broadband methods such as free space. At the range of testing frequencies, the dielectric constant of textile substrate remains unchanged [59-60].

First, the empty cavity was measured to calibrate the system. Then, placing the sample inside and using the equations (5) and , values were retrieved

$\varepsilon' = 1 + (V_c(f_c-f_s)) / (2V_sf_s)$     *(Equation 5)*

$\varepsilon'' = (V_c /4V_s) * ((Q_c-Q_s) / Q_cQ_s)$     *(Equation 6)*

where $f_c$ is the resonant frequency of the cavity when it is empty and $f_s$ is the resonant frequency when the sample is placed inside the chamber. $Q_c$ and $Q_s$ refer to the quality factor of the cavity in an empty state and a materially loaded state respectively. $V_c$ and $V_s$ represent the volume of the empty perturbation cavity and sample volume, correspondingly. We employed this technique to characterize the textile substrate in order to create a better numerical representation model. The textile sample was measured and then rotated 90 degrees and measured again in the weft direction. A dielectric constant $\varepsilon_r$ of 1.58 and a dissipation factor (*DF*) of 0.02 were obtained. The loss of bonding adhesive layer with $\varepsilon_r$ ~2 and DF ~0.02 coupled with the textile is negligible due to its relatively small thickness [61].

An essential numerical representation of an antenna is the reflection coefficient, also known as the return loss ($S_{11}$), which refers to how much of an electromagnetic wave inserted through an input port is reflected due to a discontinuity in the transmission medium. Thus it describes the ability of the antenna to radiate energy at a specific frequency range [62-63]. Another critical characteristic often listed in the antenna's specifications is the far-field pattern, or radiation pattern, that describes the intensity of the radio waves radiating from the antenna. Usually, this is a three-dimensional (3D) pattern and is often represented by spherical coordinates. Each magnitude was normalized with respect to the peak gain 2.83 dBi (decibels-isotropic), while retrieving the radiation pattern.

A test campaign was carried out at the Antenna Measurement Laboratory at Queen Mary University of London [64] to perform an electromagnetic characterization of the fabric antennas. For measuring the return losses, the antenna was connected to a vector network analyser (VNA), PNA-L Agilent N5230C. The prototypes were placed inside a mobile antenna electromagnetic compatibility (EMC)-screened anechoic chamber to examine the radiation patterns of the antenna under test (AUT) (Figure 4 (a)). The EMC chamber is equipped with two open boundary quad-ridge horn antennas operating from 400 MHz to 6 GHz (ETS-Lindgren 3164-06) and from 0.8 to 12 GHz (Satimo QH800), allowing vertical and horizontal linear polarization measurements.

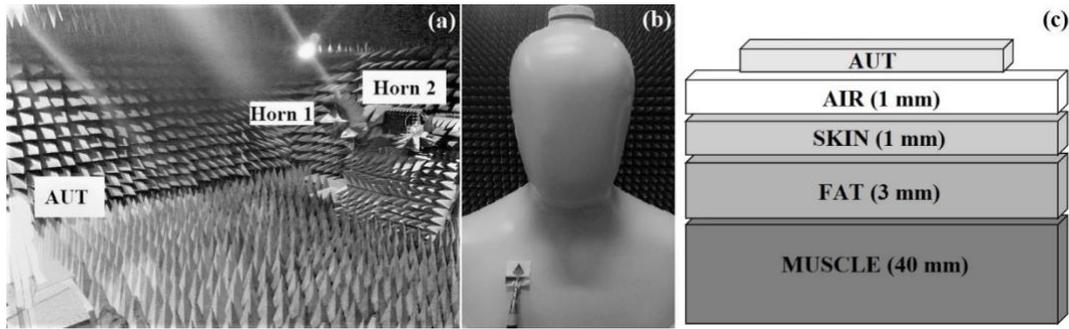

*Figure 4.* (a) EMC chamber setup used in this study (b) An image showing an AUT on a phantom (c) 4-layer phantom model used for on-body simulations

A detailed numerical analysis of the antenna performance, based on a phantom model of human body parts by using a range of $\varepsilon_r$, $\tan\delta$, $\sigma$ and $\rho$, mimicking the actual dielectric properties of the human body, was carried out. To examine the dissipative effects of the human body, a torso phantom that was filled with a liquid to replicate human tissues was used. A simplified tissue of human muscle was used as a reference because it had the highest $\varepsilon_r$ and $\sigma$, as well as occupying large volume. Different compositions of solutions were proposed in the literature [65-67]. In this study, the solution consisted of 79.7% of deionized water 0.25% of sodium chloride, 16% of Triton X-100 (polyethylene glycol mono phenyl ether), 4% of diethylene glycol butyl ether (DGBE) and 0.05% of boric acid, which filled the phantom in Figure 4(b).

The design parameters of the antenna design consist of a UWB antenna with a CPW feed and a planar inverted cone-shaped patch geometry antenna, which are shown in Figure 3. The antenna geometry was founded on the monopole disc principles [68], and the final antenna geometry was derived from the basis of optimisations. The principal equation for a quarter-wavelength monopole is

$L = \lambda/4$                                                            *(Equation 7)*

where $\lambda$ is the wavelength of the radio wave propagated and $L$ represents the physical length of the monopole antenna that determines the resonance frequency by $f_r$

$f_r = c/\lambda = c/4L$                                            *(Equation 8)*

where $c$ is the speed of light ($3 \times 10^8$ m/s).

The physical length of the monopole, $L$, determines the lowest cut-off frequency of operation. Nonetheless, this antenna design is as a concatenation of several monopoles of different sizes (in a planar inverted cone shape), so it can also support multiple resonant modes ranging from higher-order modes to harmonics from the primary resonant frequency

$f_{rn} = nc/4L$                                                 *(Equation 9)*

where $n$ is natural numbers. The operating bandwidth for the CPW feed is critically dependent on the gap between the PICA radiator patch and the CPW: in this study it is given as 0.3 mm [69]. On the contrary, to microstrip patch antennas [70], placing a conductive ground on top of the fabric provides design freedom in terms of the thickness of the fabric substrate and the required control of $\varepsilon_r$. As stated by Perez [71], the typical size of a microstrip patch ranges from $\lambda/2$ to $\lambda/3$, and the dielectric thickness, usually in the range of $0.003\lambda$ to $0.05\lambda$, corresponding to the relative constant $\varepsilon_r$, is in the 2.5 to 3.2 range. It is apparent that increasing the thickness of the dielectric buffer layer brings weight to the fabric and reduces the wearer's comfort.

A numerical analysis was carried out using CST-Microwave Studio [72] to evaluate the time-domain characteristics of the antenna structure, in which the different layers and the SMA connector were taken into account. In CST simulations, thin ML graphene layer was constructed as a conducting material which the current flows not along the surface but through the whole volume. This software calculates the development of the electromagnetic fields through time at certain spatial spots and discrete-time samples, using Maxwell's equations [73].

These antennas were also simulated on the body by using the medium of a body phantom. The phantom model consisted of a block of four layers in total, 44 mm thick, with another 1 mm air gap placed between the AUT. As depicted in Figure 4 (c), the thickness of body skin, fat and muscle layers were 1 mm, 3 mm and 40 mm respectively. This body representation with a 1 mm air gap, portrays the practical implications of wearable antennas that are not placed on the body without an isolation layer that could be another textile or a protective gel. Direct contact with the skin causes detuning of the frequency and ultimately decreases the efficiency due to the reflective behaviour. The electromagnetic properties of the different tissues: dielectric constant ($\varepsilon_r$), loss tangent ($tan\delta$), conductivity ($\sigma$), and resistivity ($\rho$), were retrieved from the reference studies [74-75] and are listed in Table 1. Even though the accurate representation of the human body on phantom models is challenging, electromagnetic properties of different body compositions retrieved at 2.45 GHz has been widely accepted in testing on-body scenarios [76-77].

*Table 1* Electromagnetic properties of human tissues at 2.45 GHZ [65]

|  | $\varepsilon_r$ | $tan\delta$ | $\sigma$ (S/m) | $\rho$ ($\Omega\cdot$m) |
|---|---|---|---|---|
| **Dry skin** | 38.007 | 0.28262 | 1.4641 | 0.683 |
| **Fat** | 5.2801 | 0.14524 | 0.1045 | 9.569 |
| **Muscle** | 52.729 | 0.24194 | 1.7388 | 0.575 |

Woven cloth exhibits strong resistance to stretching and weak stability when bent, and fabric antennas are no different. Hence, a brief simulation was carried out to reveal the bending effect on the input matching of the antenna. Previous studies have suggested that bending the *XZ*-plane (*E*-plane) had a more significant impact on the antennas' resonance length than deformation through the *YZ*-plane (*H*-plane [78-79].
As stated in fr= c/$\lambda$ = c/4L
(Equation 8, there is a proportional relationship between the antenna's resonance length and its resonance frequency. Thus, any expansion of the length through bending or stretching would have an impact on the antenna's resonance. Therefore, in this study the simulation was performed bending the antenna along the *XZ*-plane around two different cylinders with a diameter of 70 mm and 150 mm. These dimensions represent parts of the human body, e.g., the arm, the leg and the shoulder, which could be extrapolated into central angles (radians) by using the formula

*θ=W/R*                                                                                                                                            *(Equation 10)*

where *W* is the width of the antenna and *R* is the radius of the cylinder. This level of bending could be translated into degrees of angle, where the diameters of 70 mm and 150 mm correspond to 80° and 37.5° respectively.

# 3. Results and Discussion

This study attempts to reveal the on- and off-body antenna characteristics of soft graphene-based textile antennas. The return losses, $S_{11}$, both computed and measured, were investigated in free-space, on-body and under-bending settings. First, as a reference, we performed a computational analysis to predict the antenna behaviour and identify the ways in which it deviates when placed on the body. Figure 5 suggests that $S_{11}$ computational results retrieved from on- and off-body measurements demonstrated a UWB ultra-wideband behaviour rooted in the geometry of the radiator patch. The planar inverted cone-shaped antenna behaves as a concatenation of monopoles of different sizes, instead of as a single patch. The longest section of the radiating element, positioned in the middle of the antenna that was designed, was around 25 mm, which formed the longest monopole of all the concatenations. The length of this monopole determines the lowest cut-off frequency in operation, at a quarter of a wavelength for around 3 gigahertz (GHz). The dimensions of the antenna, 50 mm x 47.5 mm, are equivalent to half of the wavelength for a frequency around 3 GHz, which also addressed the lower boundary of operation. While computed on-body results cover a frequency band of over 2-10 GHz, off-body simulations exhibited a return loss less than 10 dB in a slightly narrower range, from 3-10 GHz. These slight decreases in bandwidth could be understood from the viewpoint of the dissipative medium of the body. Nevertheless, the proposed design, whether on- or off-body, covers a wide bandwidth range that meets the requirements of many wireless standards, including UWB communication [80-81].

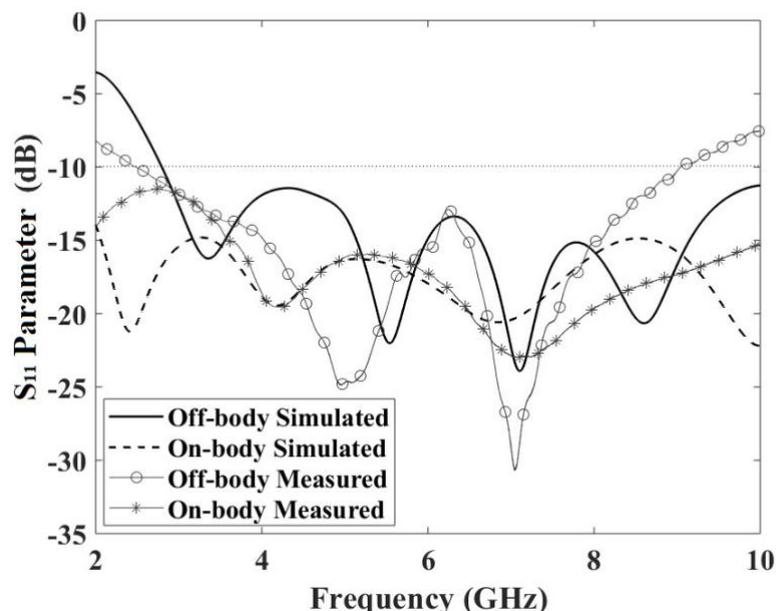

*Figure 5* Simulated vs Measured $S_{11}$ of graphene-based antennas in off- and on-body settings.

These graphene-based PICA antennas were connected to a VNA that was calibrated with a standard electronic calibration kit to the end of the coaxial cable to remove all the noise related to the cables and connectors, where the same reference was used for all the measurements. As presented in Figure 5, both measured and simulated off-body $S_{11}$ results exhibited multiple resonance frequencies at a bandwidth range of 3-9 GHz. Slight mismatches related to fabrication tolerances and unavoidable real-time losses during testing might cause such inadequacies. Nevertheless, these graphene-based soft antennas promise a 6 GHz operational bandwidth which is almost double in comparison with previous graphene-based antenna prototypes reported in the literature [37].

Textile-based antennas will be exposed to prolonged stresses of various kinds during both their use and their after-care, and thus need to be designed to conformally adhere to both irregularly shaped surfaces and the body [82]. The making strategies in wireless signal communication is of high significance for ensuring tactile comfort. We addressed three fabric coating parameters that affect the antenna performance, coupled with changes in their surface conductivity. These were: *(i)* the quality of their surface coverage; *(ii)* their vulnerability to body fluids and washing, and *(iii)* their changes in conductivity as a result of cyclic bending. As described in the introduction, due to the ease of fabrication they afford, graphene-based inks are frequently used in printing. The surface conductivity is dependent on the lateral dimensions of graphene sheets, and also on the agglomerations of these nanoflakes. Figure 6 (a) and 6 (b) demonstrate the placing of a screen-printed graphene surface onto cotton fabrics. Surface cracks could be associated with the liquid penetration process and evaporation of solvents in the inks. Nevertheless, when higher surface conductivity was attempted in the work with textiles, these nanoscopic and microscopic features caused extra challenges in achieving the level of electron conduction. Thus, increasing layers of deposition would help this mechanism in the trade-off with tactile comfort due to the additional weight and roughness. Figure 6 (c) shows ML graphene-laminated surfaces on textiles. These surface traces are caused by the lamination process; however, we observed no cracks and gaps across the textile surface. It is evident that CVD grown ML graphene sheets provided a better quality of surface coverage compared to screen-printed conductive textile surfaces. Thus, even though it involves an increase in the number of layers, it would be feasible to keep the soft, lightweight nature of the coating. Video S1 in the supplementary file shows the soft nature of ML graphene on silk weave.

As in many wearable applications in e-textiles, we take extra measures due to the susceptibility of conductive components to water and bodily fluids, such as applying hydroscopic coatings. As revealed in Figure 6 (a) and 6 (b), water molecules can propagate through the surface cracks and damage the structure itself, and make the screen printed surfaces vulnerable to washing. Thus, despite bringing additional weight, the researchers explored different coating compositions on metallic based screen printed surfaces to provide stable antenna performance with sufficient radiation efficiency after several wash cycles [83-86]. If the conductive parts are oxidized, this functional textile becomes redundant. Figure 6 (d) suggests the surface resistivity changes with respect to the washing cycles. ML graphene on cotton fabric and ML graphene on PE film both showed a substantial decrease after the first wash, reaching saturation after around 15 cycles. Figure S1 shows 'before and after' images of washed ML graphene on cotton fabrics. Figure 6 (d) revealed that even though the fabric provided additional support, the wrinkles on the surface of the fabric had a negative contribution to the surface conductivity. Thus, we observed a substantial increase in surface resistivity of ML graphene on fabric that was enforced by fabric shrinking compared to ML-graphene on PE sheets, as a result of washing instability of cotton. Figure S1 demonstrated that in longer-term use, these structures would suffer more from the risk of detaching than from oxidation or surface damage. Thus, strong interfacial bonding would also determine the life of these fabric antennas.

Figure 7 (a) reveals the effect of 5 and 50 wash cycles on computational $S_{11}$ results of graphene-based antennas and how the drastic change in conductivity impacted the gain. While the operational bandwidth remained approximately constant in every three cases, $S_{11}$ of graphene-based antennas reduced and slightly shifted towards lower frequencies. After 50 wash cycles, the antenna gain showed a -2.3 dB decrease. All these deviations in performance opened up another discussion in wearable antennas, apart from the stability of the structure that we might look into tuneable monitoring systems. Figure S3 presents the on-body representation of washed graphene-based antennas by depicting the $S_{11}$.

It is intriguing how these graphene-based antennas suffer from the distortions associated with body movement. Ergoktas *et al.* [39] performed cyclic bending and compression tests on cotton/ML graphene/PE samples and measured the sheet resistance change over time in order to understand the stability of the ML graphene on cotton samples for wearable applications. The results revealed that the resistivity of ML graphene attached to cotton samples exhibited a steady profile and was not affected by cyclic bending. The smooth and interconnected graphene sheets might have a positive contribution to bending stability. However, it is not yet understood to what extent antennas could compensate for these slight changes in resistivity.

Building on the knowledge of bending endurance, we examined the conformity of graphene-based antennas to assess the effects of bending, as presented in Figure S2. To this effect, these antennas were placed on two foam cylinders that promote a bending effect in a similar degree of conformity to that of human limbs. The foam cylinders that have a dielectric constant of 1 close to the air were used to measure the bending effect alone. Figure 7 (b) demonstrates the simulated return loss as a function of substrate curvature. Figure 7 (b) clearly points to the presence of a disseminative medium, the body. Even though there is a strong correlation between simulated and measured return loss, a slight frequency of detuning was noted in each bending ratio.

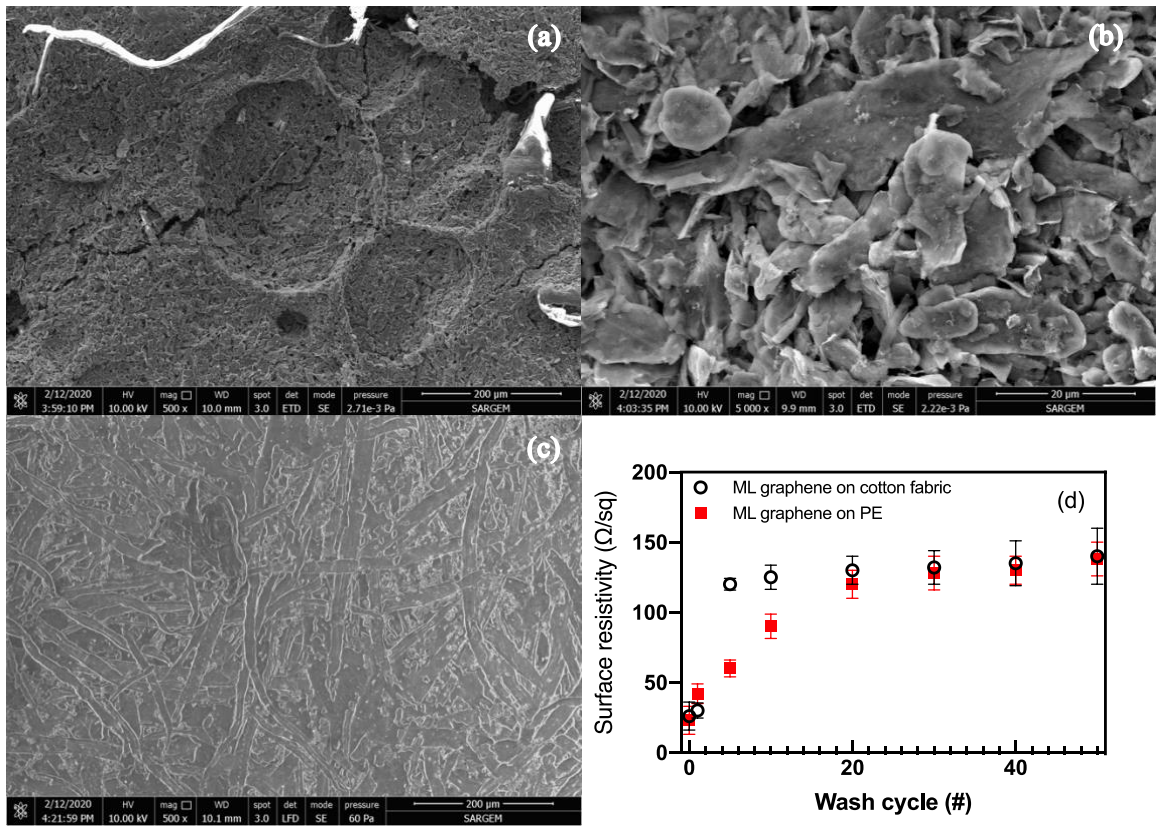

*Figure 6* SEM micrographs of graphene-based inks screen-printed on cotton fabric (a) at 500x and (b) 5kx magnification (c) SEM micrographs of ML graphene laminated on cotton fabric at 500x magnification and (d) Surface resistivity changes of ML graphene on cotton and ML graphene on PE sheets according to wash cycles

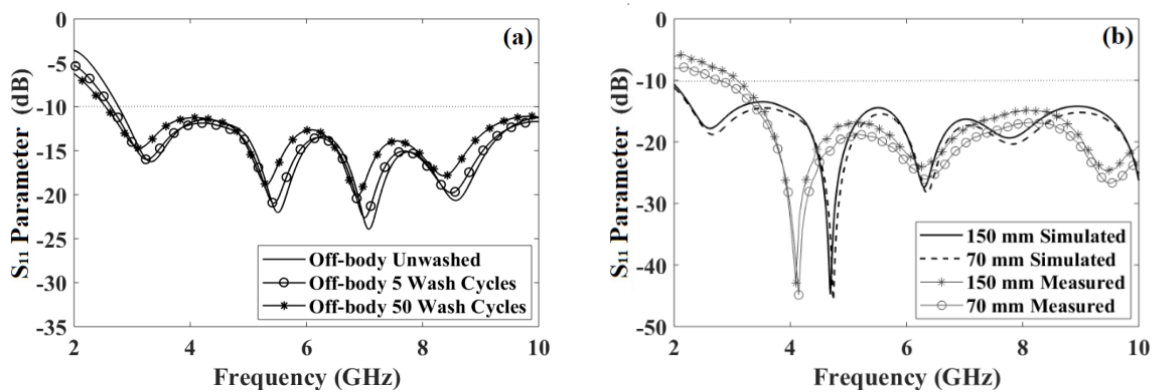

*Figure 7* (a) Simulated $S_{11}$ of unwashed and washed graphene-based antennas after 5 and 50 cycles in free space (b) Simulated vs Measured $S_{11}$ of graphene-based antennas under bending

Return loss measurements, in combination with the radiation pattern and gain measurements, show the effectiveness of graphene-based antennas in wireless personal and body area networks, targeting long-range communications such as cellular networks, wireless LAN and personal area networks. The radiation pattern taken inside an anechoic chamber presented in Figure 8 was measured in order to evaluate the radiation properties of the antenna in free space and on the phantom. Figure 8 shows the CST-simulated radiation patterns (black) of the antenna, along with the measurements (light grey) in free-space settings. Three main frequencies along the operation frequency band were selected as 3, 5 and 7 GHz. In each frequency, electric field *E*-plane (φ = 90˚) and magnetic field *H*-plane cuts (φ = 0˚) were shown. The calculation of radiation patterns is provided in detail in the supporting information. The shape of the radiation pattern is fairly consistent for the three sets of frequencies – 3, 5 and 7 GHz – and correlated well with the simulations with minimum frequency detuning. The antenna exhibited an omnidirectional pattern; this was expected, due to its CPW design. At higher frequencies (5 and 7 GHz i.e.), we observed slightly distorted omnidirectional patterning that could be related to the antenna's ability to resonate or operate in an ultra-wideband setting while the design and its dimensions were optimized for use at 3 GHz.

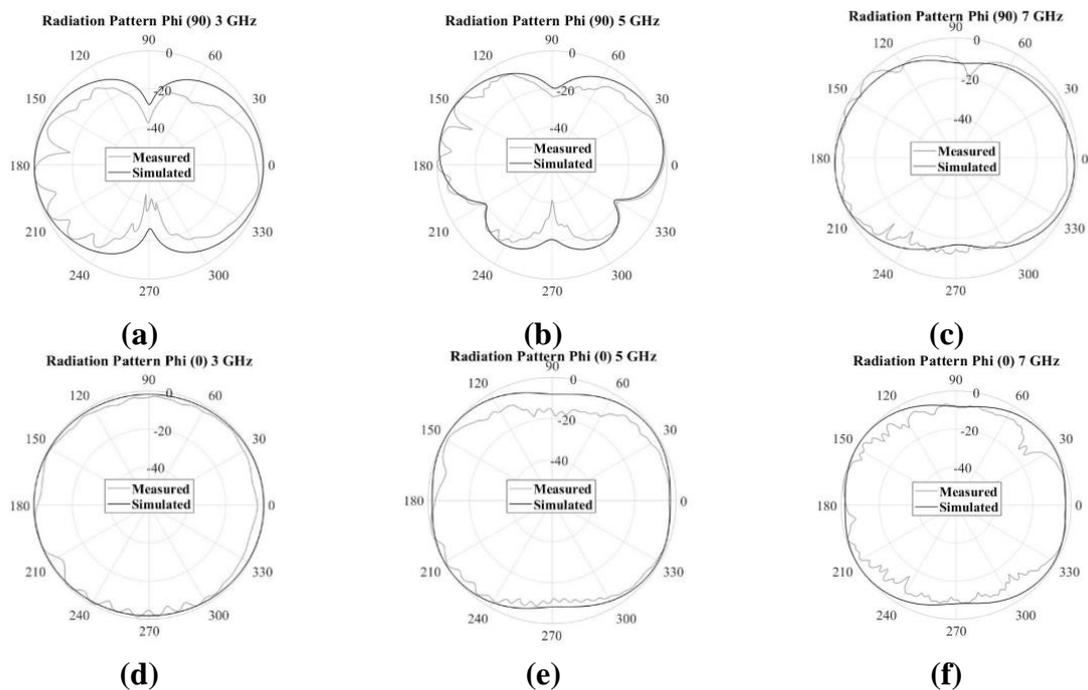

*Figure 8* Measured vs simulated radiation pattern of the graphene-based textile antenna in free space at: *E*-plane cut, at φ = 90˚: (a) 3 GHz, (b) 5 GHz and (c) 7 GHz; *H*-plane cut, at φ = 0˚: (d) 3 GHz, (e) 5 GHz and (f) 7 GHz.

The antenna was placed on the phantom's chest, where the antenna suffers less physical stress [77,81-82], during the on-body measurements. The radiation pattern of the antenna was also measured, in order to evaluate the radiation properties of the antenna on a phantom with no arms and legs [87-88]. Figure 9 (a-f) presents the measured and simulated radiation patterns for the antenna at three different frequencies, 3 GHz, 5 GHz and 7 GHz, on the phantom. The results suggest that the on-body measurements agree significantly with the simulated results. However, due to the absence of shielding that isolates the antenna from the dissipative medium of the human body, the antenna radiation performance drastically differed from the free-space measurements including the front-to-back ratio and main radiation beam properties. In fact, the human body absorbed the reflected waves that accompanied radiation at the front. From Figure 9 (a-f) it can be observed that the torso phantom absorbs more energy than the numerical phantom model. This deviation in behaviour could be due to the complexity of reproducing the human body and all its properties. Antenna design itself with a directivity function pointing towards the centre of the curvature enhanced the intensity of surface waves and increased the radiation into the body.

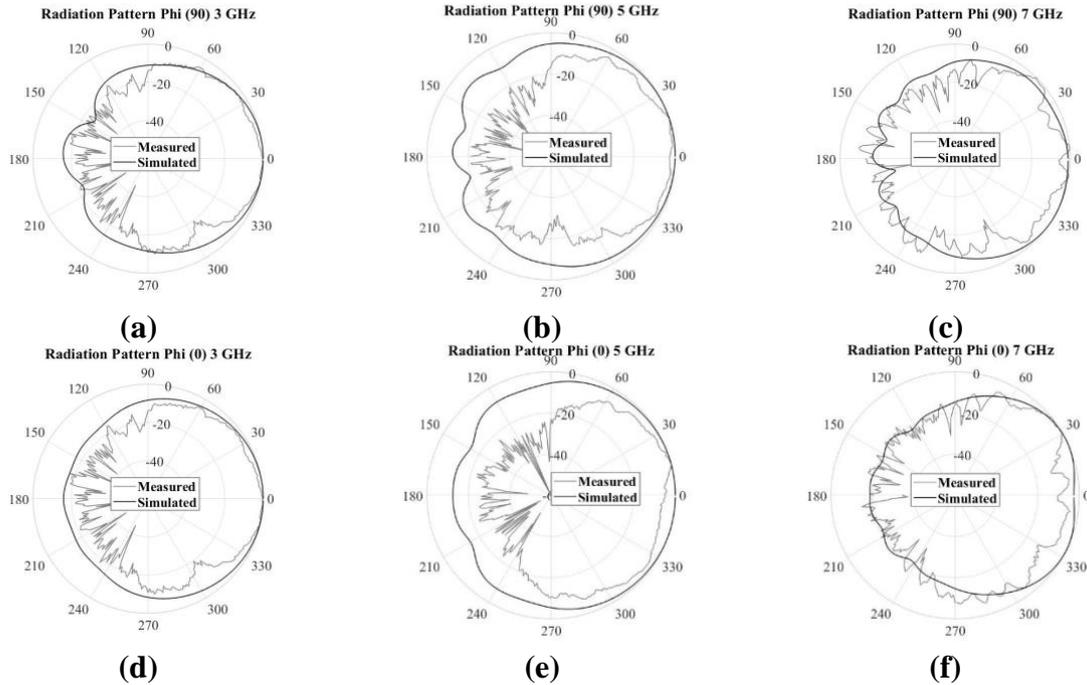

*Figure 9* Measured vs simulated radiation pattern of the proposed graphene-based antenna on the phantom at: *E*-plane cut, at φ = 90˚: (a) 3 GHz, (b) 5 GHz and (c) 7 GHz; *H*-plane cut, at φ = 0˚: (d) 3 GHz, (e) 5 GHz and (f) 7 GHz.

## 4. Conclusions

Beyond the successful design of radio communication interfaces and antennas, ensuring user-friendliness and the seamless integration of technology into the fabric is a still major challenge. Today, in addition to addressing well-known mechanical requirements such as lightness, compactness and flexibility, researchers are exploring the invisible and unobtrusive integration of smart components into garments. The emphasis on 21$_{st}$-century sustainability issues in textiles has given us a reason to question the possible implications of e-textiles that rely on the use of metals. Thus we are seeking a deep understanding of smart ways in which 2D carbon materials can be used in e-textiles as an alternative to metals. We introduced CVD grown ML graphene sheets onto a cellulosic textile substrate, in which the designed antenna consists of a CPW-fed planar inverted cone-shaped patch geometry, which is tuned to ensure the wearer's comfort. The results presented in this paper show a high level of agreement between the numerical estimation and capture-the measured data. The results demonstrate an operational bandwidth of at least 6 GHz (3–9 GHz) both in free-space and on-phantom settings. The graphene-based textile antennas were tested by addressing issues of conformity. The results under bending revealed only a slight frequency detuning in the performance that could be negligible owing to the antenna's wide bandwidth operation. Numerical and experimental analyses were carried out in order to determine the effect of the presence of the human body on the antenna return loss and radiation in comparison to the free-space scenario. While an omnidirectional radiation pattern was observed in the free-space scenario due to its co-planar structure, on-body measurements pointed to the antenna's backward radiation that was absorbed by the human phantom, leading to an enhanced back to front ratio. In summary, the proposed design that was tuned for textiles offers significant structural advantages, such as lightness, washability and drapability, along with performance advantages that include a wide radiation pattern for maximum coverage across the body, minimum radiation towards the body due to being placed full ground plane and stability subject to bending.

**Supporting Information is available.**

# Acknowledgements

This project has received funding from the European Union's Horizon 2020 research and innovation programme under grant agreement No 796640.